\documentclass[12pt,preprint]{emulateapj}
\usepackage{amssymb}

\shortauthors{A.~J. Cuesta et al.}
\shorttitle{Dark Matter decay and annihilation in the Local Universe: CLUES from \emph{Fermi}}

\begin{document}

\title{Dark Matter decay and annihilation in the Local Universe: CLUES from \emph{Fermi}}
\author{A.~J.~Cuesta\altaffilmark{1}, T.~E.~Jeltema\altaffilmark{2}, F.~Zandanel\altaffilmark{1}, S.~Profumo\altaffilmark{3,4}, F.~Prada\altaffilmark{1,5}, G.~Yepes\altaffilmark{6}, A.~Klypin\altaffilmark{7}, Y.~Hoffman\altaffilmark{8}, S.~Gottl\"ober\altaffilmark{9}, J.~Primack\altaffilmark{3}, M.~A.~S\'anchez-Conde\altaffilmark{10,11} and C.~Pfrommer\altaffilmark{12}}
\altaffiltext{1} {Instituto de Astrof{\'\i}sica de Andaluc{\'\i}a (CSIC), E-18008 Granada, Spain. E-mail: ajcv@iaa.es (AJC), fabio@iaa.es (FZ)}
\altaffiltext{2} {UCO/Lick Observatories, Santa Cruz, CA 95064, USA, E-mail: tesla@ucolick.org (TEJ)}
\altaffiltext{3} {Department of Physics, University of California, Santa Cruz, CA 95064, USA}
\altaffiltext{4} {Santa Cruz Institute for Particle Physics, University of California, Santa Cruz, CA 95064, USA}
\altaffiltext{5} {Visiting research physicist at the Santa Cruz Institute for Particle Physics, University of California, Santa Cruz, CA 95064, USA}
\altaffiltext{6} {Universidad Aut\'onoma de Madrid, Grupo de Astrof{\'\i}sica, E-28049 Madrid, Spain}
\altaffiltext{7} {Department of Astronomy, New Mexico State University, Las Cruces, NM 88003-0001, USA}
\altaffiltext{8} {Racah Institute of Physics, Hebrew University, Jerusalem 91904, Israel}
\altaffiltext{9} {Astrophysical Institute Potsdam, 14482 Potsdam, Germany}
\altaffiltext{10} {Instituto de Astrof{\'\i}sica de Canarias, E-38200 La Laguna, Tenerife, Spain}
\altaffiltext{11} {Departamento de Astrof{\'\i}sica, Universidad de La Laguna, E-38205 La Laguna, Tenerife, Spain}
\altaffiltext{12} {Canadian Institute for Theoretical Astrophysics, Toronto ON, M5S 3H8, Canada}

\begin{abstract}
We present all-sky simulated \emph{Fermi} maps of $\gamma$-rays from dark matter decay and annihilation in the Local Universe. The dark matter distribution is obtained from a constrained cosmological simulation of the neighboring large-scale structure provided by the CLUES project\footnote{{\tt http://clues-project.org}}. The dark matter fields of density and density squared are then taken as an input for the \emph{Fermi} observation simulation tool to predict the $\gamma$-ray photon counts that \emph{Fermi} would detect in 5 years of all-sky survey for given dark matter models. Signal-to-noise sky maps have also been obtained by adopting the current Galactic and isotropic diffuse background models released by the \emph{Fermi} collaboration. We point out the possibility for \emph{Fermi} to detect a dark matter $\gamma$-ray signal in local extragalactic structures. In particular, we conclude here that \emph{Fermi} observations of nearby clusters (e.g. Virgo and Coma) and filaments are expected to give stronger constraints on decaying dark matter compared to previous studies. As an example, we find a significant signal-to-noise ratio in dark matter models with a decay rate fitting the positron excess as measured by PAMELA. This is the first time that dark matter filaments are shown to be promising targets for indirect detection of dark matter. On the other hand, the prospects for detectability of annihilating dark matter in local extragalactic structures are less optimistic even with extreme cross-sections. We make the dark matter density and density squared maps available online at {\tt http://www.clues-project.org/articles/darkmattermaps.html}.

\end{abstract}  


\section{Introduction}
\label{sec:intro}

A large amount of astrophysical evidence suggests that most of the Universe's matter content is in the form of cold dark matter (DM). However, the nature of DM is still one of the most important open questions in modern physics. Many different candidates have been proposed as DM constituents (see \citealt{2005PhR...405..279B} for a review on candidates, and experimental searches), yet for the time being there is no evidence in favor of any model. One of the most studied scenarios is that of weakly interacting massive particles (WIMPs), where $\gamma$-rays are generated as secondary products of WIMP decay or annihilation (e.g. \citealt{Bertonebook}). Therefore, $\gamma$-ray observations, being a complementary approach to direct searches, are a powerful tool to study the nature of DM.

At present, the Imaging Atmospheric {\v C}erenkov Telescopes (IACTs such as MAGIC, HESS and VERITAS) together with the recently launched \emph{Fermi} satellite offer great tools to search for the $\gamma$-ray emission due to DM decay or annihilation in the MeV--TeV energy range. The main instrument on board \emph{Fermi} is the Large Area Telescope (LAT), which is designed to explore the entire $\gamma$-ray sky in the 20 MeV--300 GeV energy range \citep{2009ApJ...697.1071A}. The \emph{Fermi}-LAT collaboration already reported some of their results on DM searches, with no detection of DM $\gamma$-ray emission in dwarf spheroidal galaxies \citep{2010ApJ...712..147A}, clusters \citep{fermiclus}, or spectral features (\citealt{2010JCAP...04..014A}, \citealt{2010PhRvL.104i1302A}). IACTs did not succeed in DM detection either (e.g \citealt{2008APh....29...55A}, \citealt{2009ApJ...697.1299A}, \citealt{2010ApJ...710..634A}, \citealt{2010ApJ...720.1174A}). These studies are focused mainly on DM annihilation.

Despite these negative results, the recent detection of an excess of high-energy (10--100 GeV) positrons over the standard expectation from galactic cosmic-ray models by the PAMELA experiment \citep{pamela}, triggered an interest in the possibility that these positrons originate from DM in the Milky Way. Interestingly, if DM annihilation or decay is to consistently explain the cosmic ray electron/positron data recently produced by the PAMELA, \emph{Fermi}-LAT and HESS Collaborations (\citealt{2009PhRvL.102r1101A}, \citealt{2008PhRvL.101z1104A}, \citealt{2009A&A...508..561A}, \citealt{2009PhRvL.102e1101A}, \citealt{2010PhRvL.105l1101A}), one is generically forced to consider dominantly leptonic final states to avoid over-producing anti-protons and to generate enough high-energy positrons; and a cross-section that is enhanced over its standard value by $\sim 10^3$. In turn, this is an interesting scenario for $\gamma$-ray searches, since hard final state leptons yield an unmistakable hard bremsstrahlung $\gamma$-ray spectrum \citep{2009PhRvL.103r1302P}. 

The aim of this letter is to study DM induced $\gamma$-ray emission using a constrained Local Universe cosmological simulation, as would be observed by the \emph{Fermi} satellite in an all-sky $\gamma$-ray survey for given DM models. In Section~\ref{sec:cosmosim}, we describe the cosmological simulation used to infer the DM distribution in the Local Universe. In Section~\ref{sec:fermisim}, the \emph{Fermi} observation simulations and the assumed particle physics models are described. In Section~\ref{sec:results} we present the main results from the $\gamma$-ray all-sky maps. We finally discuss the main conclusions of our work in Section~\ref{sec:conclusion}.

\section{Constrained Simulations of the Local Universe}
\label{sec:cosmosim}
In order to get a detailed description of the DM density distribution in the Local Universe, we use a high-resolution cosmological simulation box from the CLUES Project. This simulation set provides a realistic local density field which is consistent with the $\Lambda$CDM cosmology (see \citealt{2009AIPC.1178...64Y} and \citealt{2010arXiv1005.2687G} for more details). As we want to study $\gamma$-rays from large structures in the Local Universe such as nearby galaxy clusters, we choose the Box160CR simulation. This is a constrained realization with 1024$^3$ particles in a cube of 160$h^{-1}$Mpc on a side which was run using the MPI-ART cosmological code (\citealt{1997ApJS..111...73K}, \citealt{2008arXiv0803.4343G}). The initial conditions are set assuming WMAP3 cosmology (with $\Omega_m=0.24$, $\Omega_{\Lambda}=0.76$, $\Omega_b=0.042$, $h=0.73$, $\sigma_8=0.75$, and $n=0.95$) and implements the constraints from the observed density field so that it reproduces the observed matter distribution in the Local Universe on large scales at redshift $z=0$ (\citealt{1991ApJ...380L...5H}, \citealt{2003ApJ...596...19K}). The massive clusters such as Virgo, Coma and Perseus, together with the Great Attractor, are well reproduced. However, the final positions of these objects are not exactly at their observed positions, with a typical error around 5$h^{-1}$Mpc.

This cosmological simulation allows us to produce all-sky maps of the local DM density and density squared, which are proportional to $\gamma$-ray emission due to particle DM decay and annihilation, respectively (to include more distant structures, one could use a box replication technique as in \citealt{2010MNRAS.405..593Z}). We follow the method described in \citet{2008ApJ...686..262K} to compute these luminosities. The flux is proportional to $\sum_i m_p/4\pi d_i^2$ for decay and $\sum_i m_p \rho_i/4\pi d_i^2$ for annihilation, where $i$ runs from 1 to the number of particles in each pixel, $m_p$ is the mass of the simulation particle, $\rho_i$ is the density associated to the $i$-th particle, computed using the sphere which contains its 32 nearest neighbors (no smoothing kernel was used), and $d_i$ is the distance to the observer. Only particles between 5$h^{-1}$Mpc and 80$h^{-1}$Mpc from the observer, which is placed at the right distance from the Virgo cluster, are taken into account. A proper description of the density field in the innermost 5$h^{-1}Mpc$, although affected by random density fluctuations at this scale, would require a higher resolution simulation to be resolved properly (as in \citealt{libeskind}). Thus this region empty of massive large structures is not considered here. We bin these fluxes in a Cartesian grid with 3600 and 1800 pixels of galactic longitude and latitude, respectively. This corresponds to an angular resolution of roughly 0.1~deg per pixel, reproducing the best angular resolution that \emph{Fermi} has at its highest accessible energy range.

Due to the finite resolution of the simulation, we cannot resolve the very inner center of DM halos. For this reason, we correct the flux, if underestimated, in every pixel where the centers of DM halos lie (see \citealt{2008ApJ...686..262K}). We assume a NFW profile \citep{1996ApJ...462..563N} for the inner density profile of these halos and we extrapolate it up to the halo center. The scale radius $r_s$ of these halos is calculated from the virial mass--concentration relation in \citet{2008MNRAS.391.1940M}. Typical corrections do not exceed $\sim$25\% and $\sim$250\% of their original value for decay and annihilation, respectively. We note that no boost factor due to DM substructures, or any other effect (such as adiabatic contraction from baryons or Sommerfeld enhancement), is included in our analysis.

\begin{figure*}
\begin{center}
\includegraphics[width=\textwidth]{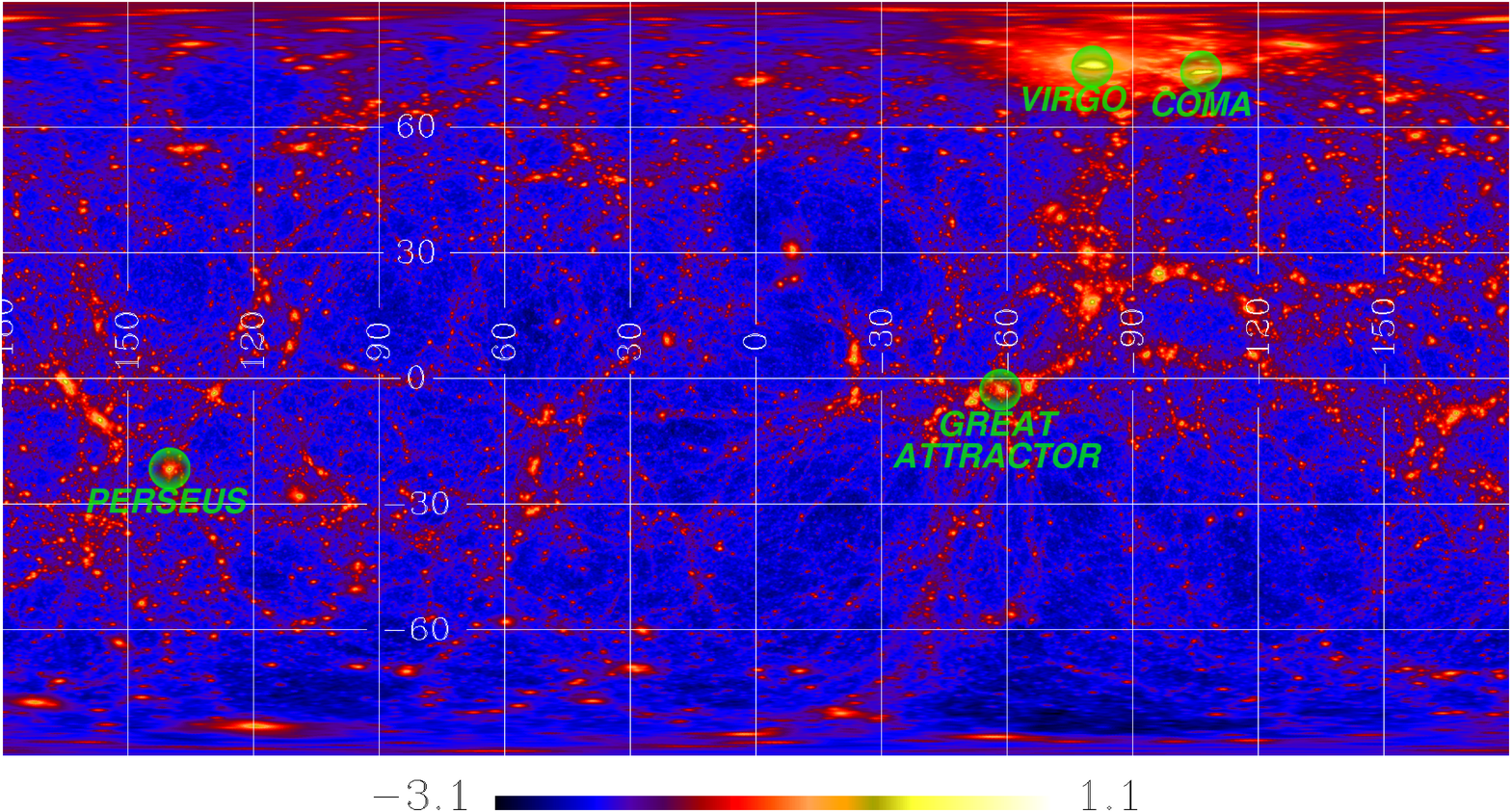}
\includegraphics[width=\textwidth]{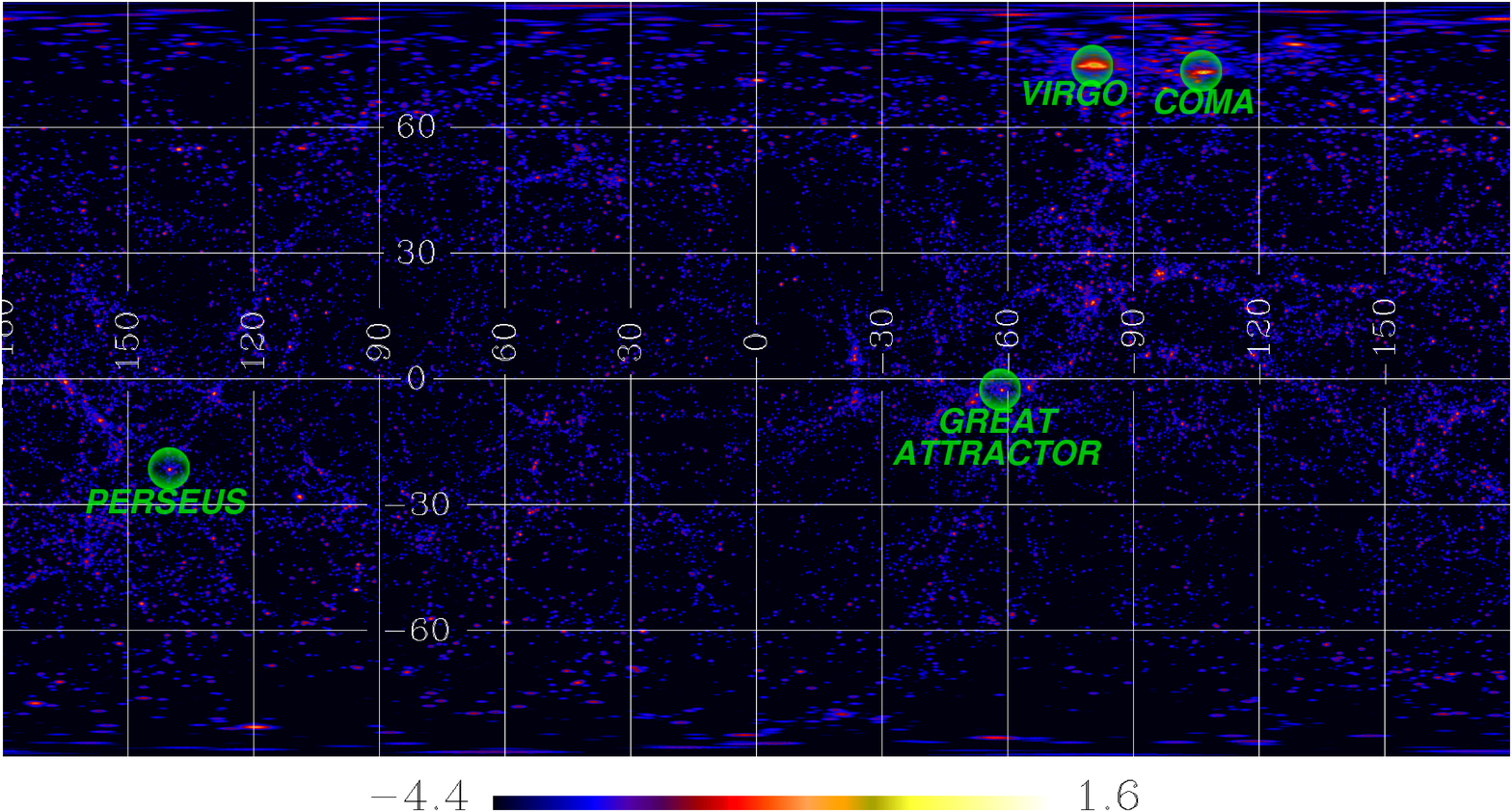}
\end{center}
\caption{DM distribution in the Local Universe constrained cosmological simulation Box160CR. These all-sky maps are Cartesian projections in Galactic coordinates. Top panel shows the density distribution, whereas the bottom panel displays the distribution of density squared. The maps are color-coded according to the $\log_{10}$ of the DM flux, and units are GeV$/c^2$ cm$^{-3}$ kpc sr$^{-1}$ for decay map and GeV$^2/c^4$ cm$^{-6}$ kpc sr$^{-1}$ for the annihilation map. Largestructures reproduced by the simulation such as Virgo, Coma, and Perseus clusters, together with the Great Attractor, are labeled. High-resolution versions of these figures are available at {\tt http://www.clues-project.org/articles/darkmattermaps.html}.}
\label{fig:density}
\end{figure*}

The resulting all-sky maps are shown in Figure~\ref{fig:density}, where known objects are highlighted. The images are done with HEALPix\footnote[1]{{\tt http://healpix.jpl.nasa.gov}}. Angular projection used here is equirectangular (plate carr{\'e}e). Maps are color-coded according to the logarithmic flux in each 0.1~deg pixel, measured in GeV$/c^2$ cm$^{-3}$ kpc sr$^{-1}$ for decay and GeV$^2/c^4$ cm$^{-6}$ kpc sr$^{-1}$ for annihilation. These maps are used as input for the \emph{Fermi}-LAT observation simulations, which we describe in the following Section.

\section{Fermi Satellite Observation Simulations}
\label{sec:fermisim}
Using the full sky DM density and density squared maps, the simulated \emph{Fermi}-LAT observations are produced using the \texttt{gtobssim} routine, part of the \emph{Fermi} Science Tools package (v9r15p2), which incorporates the \emph{Fermi}-LAT effective area and point spread function and their energy dependence. All simulations are run to generate a 5-year observation in the default scanning mode and using the current release of the LAT instrument response functions (P6\_V3\_DIFFUSE).

In the present study, we specifically adopt two examples for the $\gamma$-ray spectrum from decay or annihilation of the DM particle, chosen to be representative of more general classes of DM models. The first model features a DM particle with a mass of 1.6 TeV yielding a pair of $\mu^+\mu^-$ \citep{2009PhRvL.103c1103B}, which was shown to fit accurately the PAMELA data in \citet{2010JCAP...03..014P}. In this case, $\gamma$-ray emission is produced directly in the final state radiation (FSR) as well as through inverse Compton scattering (IC) of the high energy $e^+$ and $e^-$ produced off of cosmic microwave background photons. We include the expected contributions from both FSR and IC (see \citealt{fermiclus} for details). Given that the energy loss time scales for high-energy electrons and positrons produced by muon decays are much shorter than the diffusion time scales in the structures we consider here, we neglect diffusion, and also calculate the emission of said electrons and positrons via IC up-scattering of CMB photons. This yields a significant low-energy component, extending all the way up to energies relevant to the \emph{Fermi} telescope \citep{2009JCAP...07..020P}. We also consider a second, more conventional model, inspired by what it is expected in e.g. supersymmetric models with a bino-like lightest supersymmetric particle: a 100 GeV neutralino yielding a quark-antiquark pair (of $b$ flavor, for definiteness). The primary source of $\gamma$-rays here is the decay of neutral pions produced in the $b\bar b$ hadronization chains.

The \emph{Fermi} simulations of the $\gamma$-ray signal from DM annihilation (both to $b\bar b$ and $\mu^+\mu^-$) were normalized to a DM flux in the \emph{Fermi} energy range of $9 \times 10^{-8}$ photons cm$^{-2}$ s$^{-1}$ integrated over the full sky. This was chosen in order to obtain good statistics to compare between different extragalactic structures. For $b\bar b$, this flux corresponds to a cross-section of $10^{-23} {\rm cm}^3{\rm s}^{-1}$. This cross-section value has already been excluded by current indirect searches with \emph{Fermi}, PAMELA and HESS (see e.g. \citealt{2010NuPhB.840..284C} and \citealt{2010JCAP...03..014P}). Yet we decided to keep this extreme case for comparison purposes. For $\mu^+\mu^-$, the same total flux corresponds to an annihilation cross-section of $5.8 \times 10^{-23} {\rm cm}^3{\rm s}^{-1}$, which gives a good fit to the PAMELA positron excess. This value is only marginally excluded in \citet{fermiclus} if cluster substructures are considered, and also in \citet{2010JCAP...03..014P} if the Milky Way DM halo follows NFW.

In the case of DM decay as e.g. in supersymmetry with very weak R-parity violation induced by a dimension-6 GUT-scale operator to both $b\bar b$ and $\mu^+\mu^-$, we simulated a total \emph{Fermi} flux over the full sky of $1.5 \times 10^{-6}$ photons cm$^{-2}$ s$^{-1}$, which corresponds to a decay lifetime of $\tau \simeq 10^{26}$ s for $b\bar b$ and $\tau \simeq 3 \times 10^{26}$ s for $\mu^+\mu^-$. These lifetimes are not currently excluded by other $\gamma$-ray constraints and the latter case gives a good fit to the PAMELA measured positron fraction excess (e.g.~ \citealt{2010JCAP...03..014P}; \citealt{2010NuPhB.840..284C}; \citealt{2010JCAP...01..023C}; \citealt{2010NuPhB.831..178M}; \citealt{2010JCAP...06..027Z}).

We also include in the simulations realistic treatments of both Galactic and isotropic diffuse backgrounds. In particular, the $\gamma$-ray emission from the Galaxy is quite variable across the sky, an important consideration when comparing the expected signals from known objects. For example, structures lying at low Galactic latitudes like the Great Attractor will have much higher $\gamma$-ray backgrounds than high latitude objects like the Virgo cluster. We simulated 5-year \emph{Fermi} observations of the Galactic and isotropic diffuse backgrounds using the current background models released by the \emph{Fermi} collaboration\footnote[1]{{\tt http://fermi.gsfc.nasa.gov/ssc/data/access/lat/ BackgroundModels.html}} (\texttt{gll\_iem\_v02.fit} and \texttt{isotropic\_iem\_v02.txt}, respectively). The output background map is then used to compute signal-to-noise (S/N) all-sky maps, detailed in the next Section.

\section{Results}
\label{sec:results}

\begin{figure*}
\begin{center}
\includegraphics[width=\textwidth]{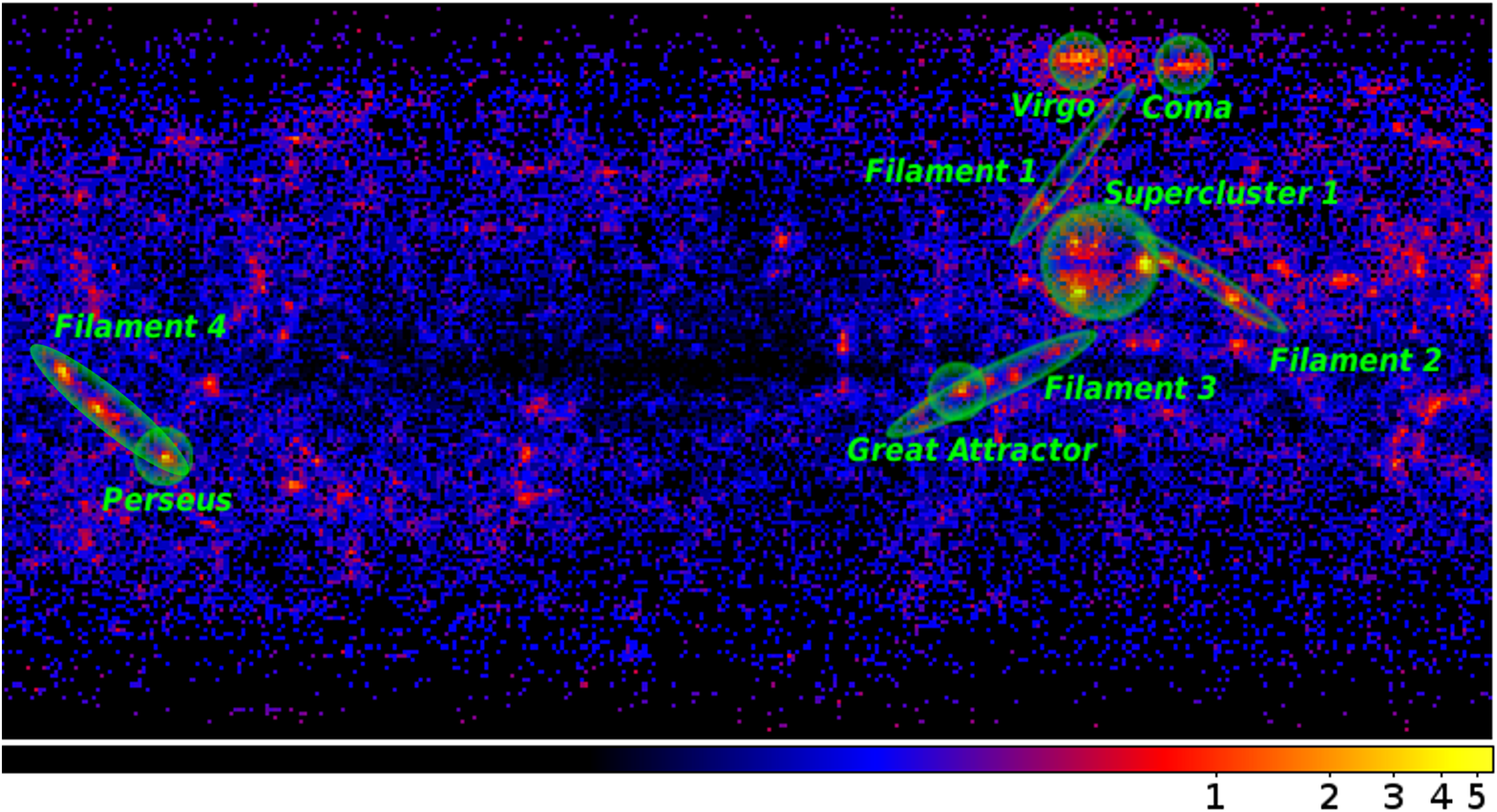}
\includegraphics[width=\textwidth]{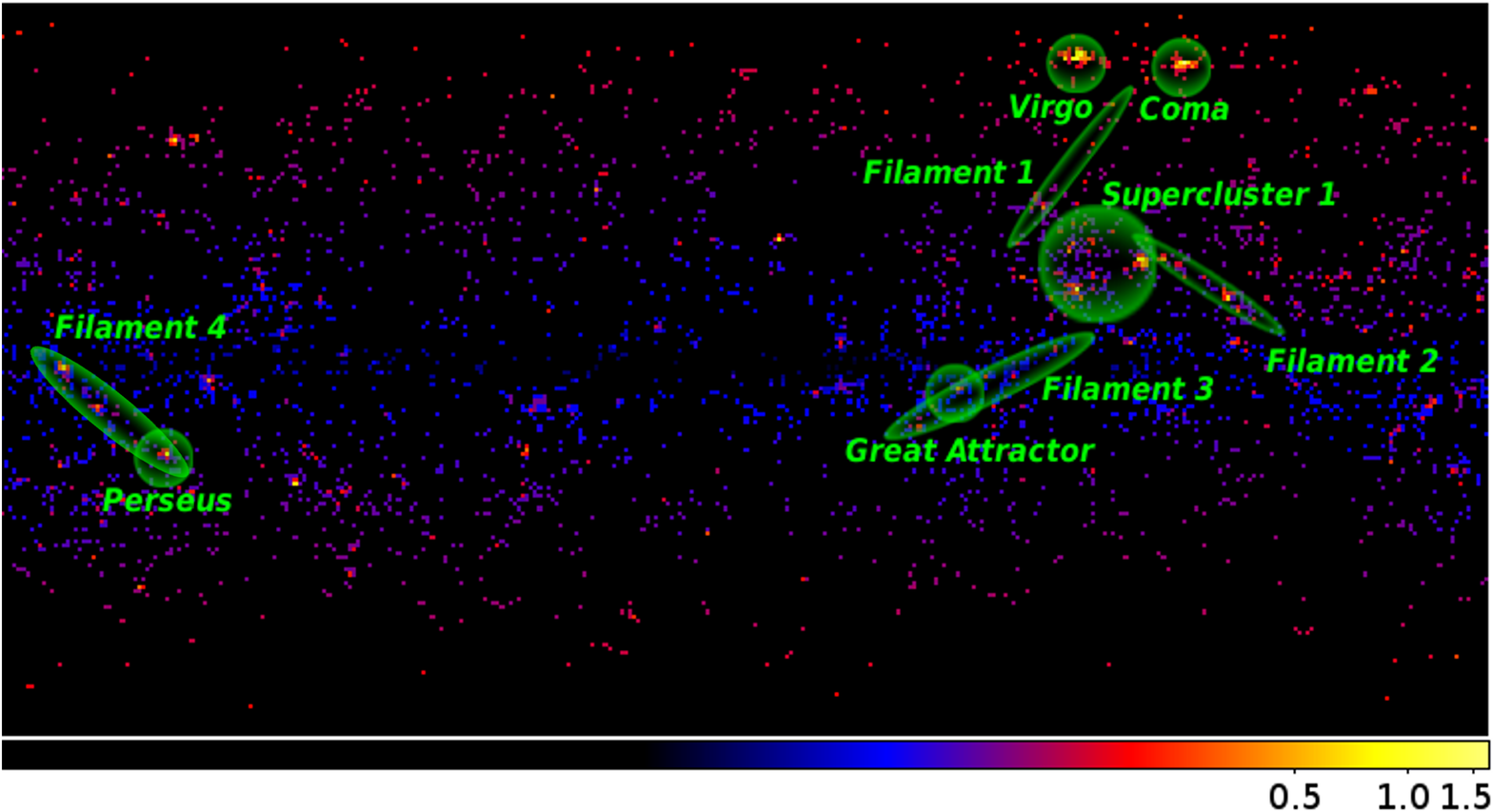}
\end{center}
\caption{S/N all-sky maps from \emph{Fermi} simulations for DM $\gamma$-rays in the energy range 100 MeV--10 GeV built from the Box160CR constrained simulation of the Local Universe. Results for DM decay (top) and annihilation (bottom) are shown for the $b\bar{b}$ channel model.}
\label{fig:maps}
\end{figure*}

\begin{deluxetable*}{lrrrrrrrrr}[!h]
\tabletypesize{\footnotesize}
\tablecolumns{10}
\tablewidth{0pt}
\tablecaption{S/N, photon counts and backgrounds in \emph{Fermi} simulations.}
\tablehead{ \colhead{Object}               & 
            \multicolumn{4}{c}{$b\,\bar{b}$ channel}  & 
            \multicolumn{4}{c}{$\mu^+\mu^-$ channel}& 
            \colhead{background}                 \\
                                    &
            \multicolumn{2}{c}{ann} &
            \multicolumn{2}{c}{dec} &
            \multicolumn{2}{c}{ann} &
            \multicolumn{2}{c}{dec} &
                                    }
\startdata
Coma 1~deg         & 2.981  & (20)   & 2.981      &  (20)  &   0.567   &   (3) & 1.078  &  (6)     & 25\\
Coma 2~deg         & 2.592  & (31)    & 4.061     &  (52)  &    0.371     &   (4)   &   1.331     &  (15)     & 112\\
Coma 5~deg         &  1.542          & (39)  &  4.976     &  (135)  &   0.203  &  (5)  & 1.542 & (39)   & 601\\
\hline
Virgo 1~deg         & 2.739 & (15) &   3.133      &  (18)  &  0.000        &   (0)   &   0.485    &  (2)     & 15\\
Virgo 2~deg         & 2.858 & (28) &  5.371       &  (61)  &  0.471       &   (4)    &  1.444    &  (13)     & 68\\
Virgo 5~deg         & 1.818           & (42)   &  7.700    &  (203)  & 0.269     &  (6)   &  2.189 &  (51)   & 492\\
\hline
Perseus 1~deg         & 1.144 & (12)  & 5.642              &  (74) &  0.493        &   (5)    &  1.841    &  (20)     & 98\\
Perseus 2~deg         & 0.700 & (14)  & 5.646             &  (128) & 0.253        &   (5)    &  1.518    &  (31)     & 386\\
Perseus 5~deg         & 0.386 & (20)  &  4.542            &  (245) &  0.116       &   (6)    &  1.187   &  (62)     & 2665\\
\hline
GAttractor 1~deg         & 0.280 & (8)   &  2.686          &  (80)  &   0.105      &   (3)    &  0.935    &  (27)     & 807\\
GAttractor 2~deg         & 0.211 & (13)  &  2.581         &  (162) &      0.049    &   (3)   &  0.696      &  (43)     & 3777\\
GAttractor 5~deg         & 0.136 & (23)  &  2.157                  &  (367) &  0.041       &   (7)   &   0.538     &  (91)     & 28572\\
\hline
Filament1, $d=65$Mpc/$h$    & 0.224 & (14) & 4.515    &  (292)   &  0.128    &   (8)  &    1.348     &  (85)     & 3891\\
\hline
Filament2, $d=40$Mpc/$h$    & 0.752 & (67) & 9.317        &  (871)  &  0.191      &   (17)  & 2.589       &  (233)    & 7869\\
\hline
Filament3, $d=65$Mpc/$h$     & 0.351 & (84) & 4.862        &  (1174) & 0.121       &   (29)  &  1.181      &  (283)    & 57127\\
\hline
Filament4, $d=55$Mpc/$h$     & 0.576 & (91) & 8.380        &  (1358) & 0.184        &   (29)  &  2.065     &  (328)    & 24904 \\
\hline
Supercluster1, $d=45$Mpc/$h$ & 0.911 & (144) &  12.598     &  (2066) & 0.254       &   (40)  &  3.334      &  (531)    & 24829\\
\enddata
\tablecomments{The S/N ratio and number of photon counts (in brackets) in the 1 GeV--10 GeV energy range for our different DM models. For cluster regions, three different radii are considered (1, 2, and 5 degrees). Filaments 1 to 4 represent elongated regions connected to these clusters which are potentially interesting due to their high S/N. Median distance of halos belonging to these filaments is indicated. Supercluster1 is a collection of massive halos which accidentally lie along the line-of-sight. Background counts from the Galactic plus extragalactic diffuse in the same regions are also listed. Note that the annihilation to $b\,\bar{b}$ case is shown for comparison purposes only.}
\label{tab:table}
\end{deluxetable*}

In Figure~\ref{fig:maps} we show our main result: the S/N prediction for the extragalactic $\gamma$-ray emission in the 100 MeV--10 GeV energy range from annihilation and decay of DM in the Local Universe, as it would be seen by the \emph{Fermi} satellite after 5-years of observations. This is the first time that a constrained cosmological simulation is used to generate maps that are consistent with both the currently accepted cosmology and the observed Local Universe. These maps assume a particle mass of 100~GeV which annihilates or decays through the $b \bar b$ channel. Maps for the DM model yielding $\mu^+\mu^-$ are similar but present lower S/N. Pixels are binned in squares of 1~deg which matches the \emph{Fermi}-LAT PSF at around 1~GeV as well as the typical angular size in the sky of nearby clusters. In order to make a quantitative analysis of DM detectability in large nearby structures, including cluster and filament regions, we computed in Table~\ref{tab:table} the photon number counts and S/N from annihilation and decay in the 1 GeV--10 GeV energy range, for both the $b \bar b$ and $\mu^+\mu^-$ channels. The S/N is defined as the signal over the square root of the signal plus the background $\gamma$-ray emission. This choice of energy range maximizes the S/N ratio as compared to the 100 MeV--10 GeV range, as the \emph{Fermi}-LAT sensitivity is significantly worse at lower energies. 

In the case of DM decay, we find that nearby clusters and filamentary regions could be detected for decay lifetimes longer than those currently ruled out by other $\gamma$-ray constraints as discussed above. This shows that extragalactic structures are excellent targets to search for a signal or to place constraints on DM decay models, including those fitting the PAMELA positron data. It is important to note that according to Table~\ref{tab:table} the most promising clusters for DM studies are high galactic latitude objects, like Virgo and Coma. The former, however, has not been used in the recent \emph{Fermi} search for DM in clusters, due to the presence of M87, which is a powerful $\gamma$-ray source, which makes the extraction of a signal or a limit much more complicated than for other clusters. Moreover, we find that S/N is not very sensitive to the area of the region under analysis, provided that the aperture radius is no more than few degrees, where the signal saturates and therefore the background noise makes the S/N decrease.

We also highlight that, in the case of DM decay, the filamentary structure of the cosmic web constitutes an interesting target for DM searches. To our knowledge, this is the first time that filaments have been considered as targets for DM searches. In this case the $\gamma$-ray luminosity is just proportional to the enclosed mass, whereas this is only approximately true for annihilation \citep{2009PhRvL.103r1302P}. This means that massive extragalactic objects offer the best chance for detection (see Table~\ref{tab:table}). Large filaments of DM match and even exceed the values of S/N as compared to those in large clusters, although caution should be taken regarding the exact orientation in the sky of these filaments as we find some variation in smaller volume constrained simulations. Superclusters such as the region marked in Figure~\ref{fig:maps} show even more significant values. Hence, these features of the Large-Scale Structure of the Universe may prove to be a very promising novel way to detect decaying DM with \emph{Fermi}.

\section{Discussion and Conclusion}
\label{sec:conclusion}

In this letter we have presented simulated \emph{Fermi} maps of extragalactic $\gamma$-rays coming from DM annihilation and decay in the Local Universe. The DM distribution is taken from a constrained cosmological simulation of the Local Universe Box160CR by the CLUES project, and it is available online at the following URL: {\tt http://www.clues-project.org/articles/ darkmattermaps.html}. This distribution is then taken as an input for the \emph{Fermi} observation simulation to obtain the all-sky distribution of $\gamma$-ray photon counts that would be measured by \emph{Fermi} in a 5-year survey. Galactic and isotropic $\gamma$-ray diffuse backgrounds are also taken into account. This allows us to get S/N all-sky maps to estimate the possibility of detection of $\gamma$-rays from DM in such a survey. We adopted two different particle physics models: a DM particle annihilating (decaying) primarily to a $b\bar b$ final state, and a DM model that gives a good fit to the local positron fraction measured by PAMELA and the total electron spectrum measured by \emph{Fermi} with annihilation (decay) to a $\mu^+\mu^-$ final state. 

In the case of DM decay, we find for these models that large clusters with high galactic latitudes, like Virgo and Coma, offer the best chance to be detected together with filamentary regions and large superclusters. Besides, we find that the S/N is not a strong function of the area of the analyzed region, which allows for considering large apertures without significant penalty in the results by masking away $\gamma$-ray point-sources. This is an important result as the \emph{Fermi} collaboration and other authors have started to severely constrain models of annihilating DM (\citealt{fermiclus}, \citealt{2010JCAP...04..014A}, \citealt{2010PhRvL.104i1302A}, \citealt{2010ApJ...712..147A}) while decaying DM has comparatively received considerably smaller attention (however, see \citealt{2010JCAP...07..008H}). Currently available $\gamma$-ray observations provide less stringent constraints in this case, and from a theoretical standpoint, decaying DM is a generic prediction of many theories beyond the Standard Model of particle physics, for both neutralino DM and models that explain the PAMELA positron excess. 

We do not find any strong arguments in favor of a possible \emph{Fermi} detection of extragalactic $\gamma$-rays induced by DM annihilation. However, we cannot exclude completely this possibility since we are not considering possible boost factors from DM substructures, adiabatic compression, and Sommerfeld effect that may significantly enhance the final $\gamma$-ray emission. Besides, the predicted signals will be enhanced due to recent determinations of the cosmological parameter $\sigma_8$ suggesting a higher value than the one assumed here, which is consistent with WMAP3 cosmology.

These conclusions should be complemented by an analysis of the galactic components, mainly from DM subhalos like those hosting the DM-rich dwarf spheroidal galaxies around the Milky Way. This has been recently addressed in \citet{2010ApJ...718..899A} based on the analysis of a high-resolution simulation of a galactic DM halo, as in previous papers by \citet{2008ApJ...686..262K} and \citet{2008Natur.456...73S}. We note that the presence of any galactic foregrounds not modeled here has a potential effect on the significance of our predictions, although only the Galactic Center and massive subhalos have been shown in these papers to be relevant. Moreover, the annihilation or decay of DM in Galactic subhalos will produce $\gamma$-ray photons similar to those from DM in local extragalactic structures and thus if these happened to be coincident in the sky it would only enhance the signal. Nevertheless, a spectral confirmation of the potential DM signal is necessary to validate any claim of detection. On the other hand, these results on DM search will benefit from additional hints from the study of the angular power spectrum of the $\gamma$-ray flux (e.g. \citealt{2009PhRvD..80b3518F}, \citealt{2009arXiv0912.1854H}, see also \citealt{2010MNRAS.405..593Z} for a similar approach to that presented here, extended to more distant contributions). An analysis of the anisotropies in the extragalactic diffuse radiation from DM using constrained simulations will be presented elsewhere.

Concluding, we find that \emph{Fermi} will be able to place strong constraints on the nature of dark matter by studying extragalactic objects, in particular for decay. The theoretical predictions from constrained simulations should provide the astroparticle community with the most interesting prospects for the detection of the elusive DM particle.\\

We thank support of Spanish MICINN’s Consolider-Ingenio 2010 Programme MULTIDARK CSD2009-00064 and ASTROMADRID (S2009/ESP-146). A.J.C. thanks support from MEC Spanish grant FPU AP2005-1826. GY acknowledges support of MICINN research grants FPA2009-08958 and AYA2009-13875-C03-02. SP is supported by NASA, DoE and NSF. YH is supported by Israel Science Foundation (13/08). SG acknowledges support of DAAD through PPP program. BOX160CR simulation has been performed at Leibniz Rechenzentrum Munich (LRZ). \emph{Fermi}-LAT simulation tools were provided by the \emph{Fermi}-LAT collaboration and the \emph{Fermi} Science Support Center.

\newpage
\section{Appendix: Erratum to the paper}

The authors corrected a mistake in the computation of the area assigned to each pixel in the signal maps, which was missing a factor of $0.5\pi\cos(b)$. While this error is not very important at mid-latitudes, the area of the pixels at the Galactic equator is 57\% larger than previously estimated (hence lowering the signal), and is a factor of 2.5 smaller at the position of Virgo $(b \simeq 75^{\circ})$, increasing the previously estimated signal arbitrarily as we approach to the Galactic poles. The signal-to-noise values of Table~1, as well as the maps in Figure~2 of the original paper are affected, and we report above the corrected version of both the Table and the Figure. Note that the dark matter density and density squared maps of Figure~1 of the original paper are not affected by this error. The conclusions in the original paper remain valid and the values of the signal-to-noise for the high-latitude objects we focused in the paper (mainly Virgo and Coma) are now even stronger. \\

We have made the updated FITS files for the dark matter density and density squared maps available online at: {\tt http://www.clues-project.org/articles/darkmattermaps.html}.\\

\begin{figure*}
\begin{center}
\includegraphics[width=0.45\textwidth]{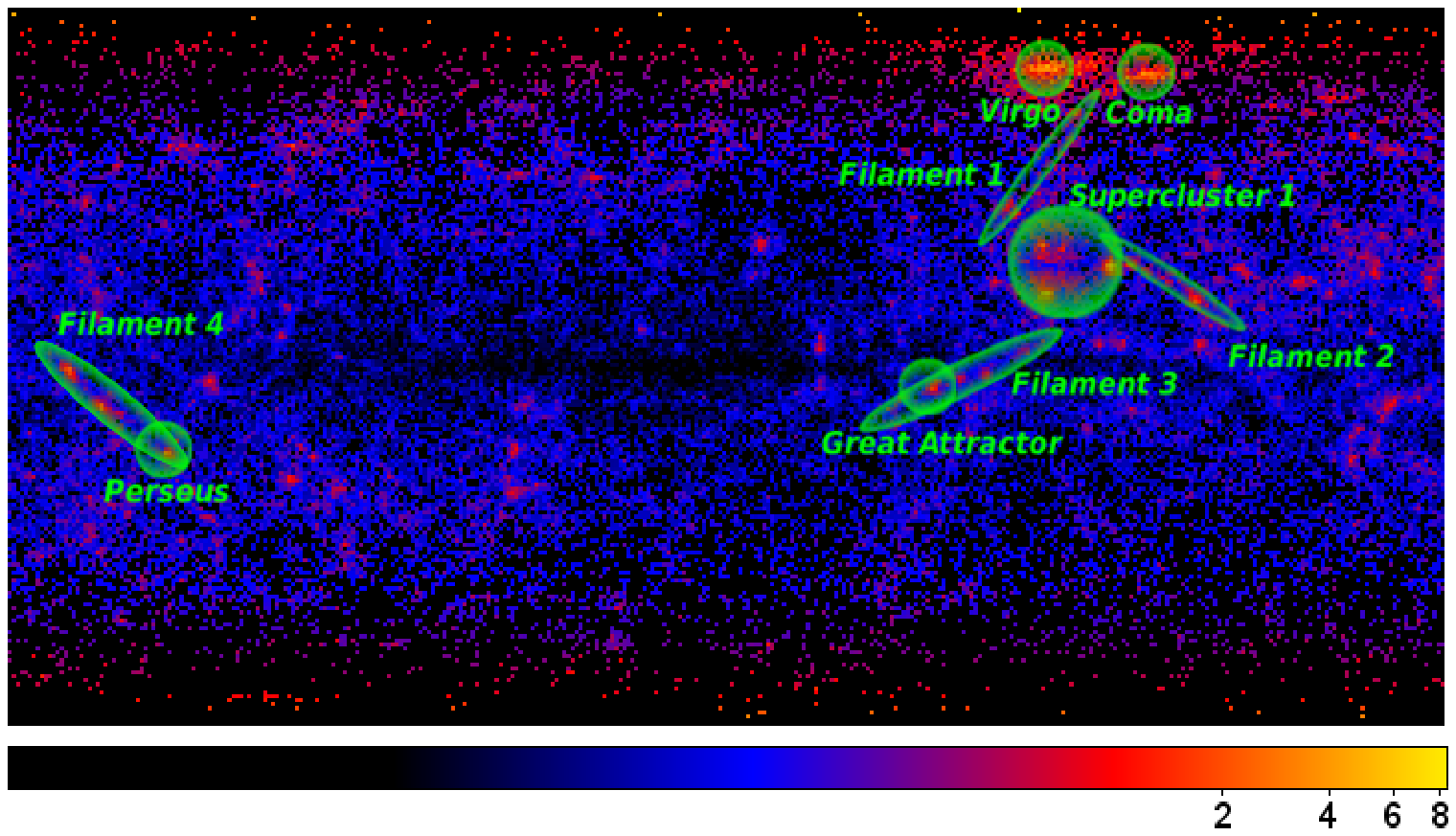}
\includegraphics[width=0.45\textwidth]{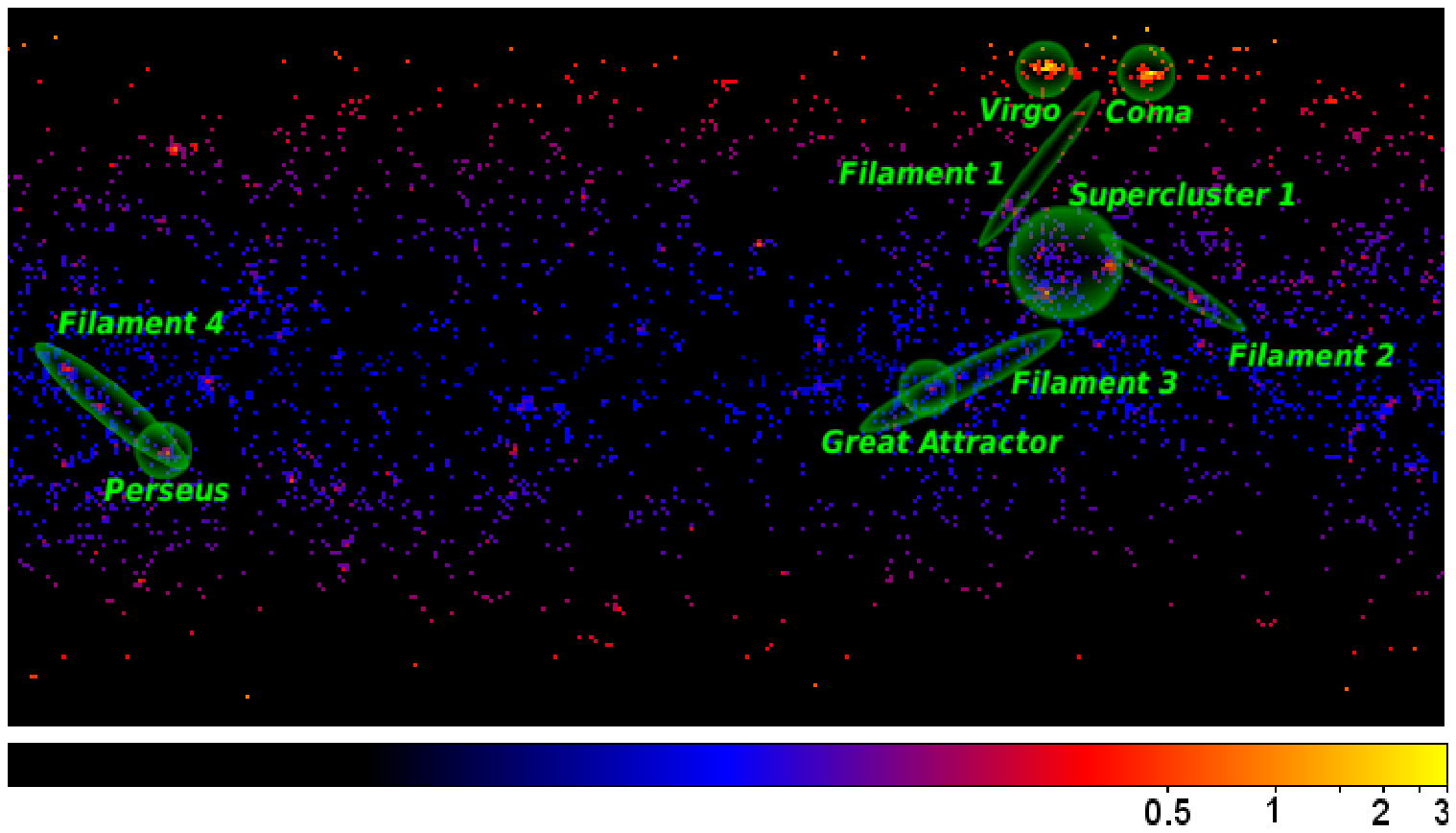}
\end{center}
\caption{S/N all-sky maps from \emph{Fermi} simulations for DM $\gamma$-rays
in the energy range 100 MeV--10 GeV built from the Box160CR constrained
simulation of the Local Universe. Results for DM decay (left) and annihilation
(right) are shown for the $b\bar{b}$ channel model.}
\label{fig:maps}
\end{figure*}

\begin{deluxetable*}{lrrrrrrrrr}[!b]
\tabletypesize{\footnotesize}
\tablecolumns{10}
\tablewidth{0pt}
\tablecaption{S/N, photon counts and backgrounds in \emph{Fermi} simulations.}
\tablehead{ \colhead{Object}               &
            \multicolumn{4}{c}{$b\,\bar{b}$ channel}  &
            \multicolumn{4}{c}{$\mu^+\mu^-$ channel}&
            \colhead{background}                 \\
                                    &
            \multicolumn{2}{c}{ann} &
            \multicolumn{2}{c}{dec} &
            \multicolumn{2}{c}{ann} &
            \multicolumn{2}{c}{dec} &
                                    }
\startdata
Coma 1~deg         & 5.297  & (44)   & 5.297      &  (44)  &   1.237   &   (7)
& 2.109  &  (13)     & 25\\     
Coma 2~deg         & 5.068  & (68)   & 7.583      &  (114) &   0.818   &
(9)   &   2.741     &  (33)     & 112\\
Coma 5~deg         &  3.245 & (85)  & 10.078      &  (303)  &  0.445   &  (11)
      & 3.245 & (85)   & 601\\
\hline
Virgo 1~deg         & 5.041 & (36) &  5.646       &  (43)  &  0.000        &
(0)    &   1.118    &  (5)     & 15\\
Virgo 2~deg         & 5.831 & (68) &  10.025      &  (147)  & 1.132       &
(10)    &  3.116    &  (31)     & 68\\
Virgo 5~deg         & 4.185           & (102)   &  15.588    &  (488)  & 0.666
&  (15)   &  5.068 &  (126)   & 492\\
\hline
Perseus 1~deg         & 0.777 & (8)  & 4.178              &  (51) &  0.298
&   (3)    &  1.323    &  (14)     & 98\\
Perseus 2~deg         & 0.503 & (10)  & 4.042             &  (88) & 0.152
&   (3)    &  1.041    &  (21)     & 386\\
Perseus 5~deg         & 0.271 & (14)  &  3.156            &  (168) &  0.077
&   (4)    &  0.807   &  (42)     & 2665\\
\hline
GAttractor 1~deg         & 0.175 & (5)   &  1.741          &  (51)  &   0.070
&   (2)    &  0.592    &  (17)     & 807\\
GAttractor 2~deg         & 0.130 & (8)  &  1.654         &  (103) &
0.033    &   (2)   &  0.438      &  (27)     & 3777\\
GAttractor 5~deg         & 0.089 & (15)  &  1.379                  &  (234) &
0.024       &   (4)   &   0.343     &  (58)     & 28572\\
\hline
Filament1, $d=65$Mpc/$h$    & 0.224 & (14) & 4.485    &  (290)   &  0.112    &
(7)  &    1.379     &  (87)     & 3891\\
\hline
Filament2, $d=40$Mpc/$h$    & 0.517 & (46) & 6.541        &  (602)  &  0.135
&   (12)  & 1.797       &  (161)    & 7869\\
\hline
Filament3, $d=65$Mpc/$h$     & 0.226 & (54) & 3.117        &  (750) & 0.079
&   (19)  &  0.756      &  (181)    & 57127\\
\hline
Filament4, $d=55$Mpc/$h$     & 0.380 & (60) & 5.486        &  (881) & 0.120
&   (19)  &  1.338     &  (212)    & 24904 \\
\hline
Supercluster1, $d=45$Mpc/$h$ & 0.640 & (101) &  8.915     &  (1445) & 0.177
&   (28)  &  2.343      &  (372)    & 24829\\
\enddata
\tablecomments{The S/N ratio and number of photon counts (in brackets) in the
1 GeV--10 GeV energy range for our different DM models. For cluster regions,
three different radii are considered (1, 2, and 5 degrees). Filaments 1 to 4
represent elongated regions connected to these clusters which are potentially
interesting due to their high S/N. Median distance of halos belonging to these
filaments is indicated. Supercluster1 is a collection of massive halos which
accidentally lie along the line-of-sight. Background counts from the Galactic
plus extragalactic diffuse in the same regions are also listed. Note that the
annihilation to $b\,\bar{b}$ case is shown for comparison purposes only.}
\label{tab:table}
\end{deluxetable*}

AJC would like to thank the corresponding authors of {\tt http://arxiv.org/abs/1110.3305} (Gomez-Vargas et al., submitted to JCAP) for realizing this error when doing the analysis of their signal-to-noise maps.
\end{document}